\documentclass[sigconf]{acmart}

\usepackage{bbm}

\usepackage{bm}
\usepackage{algorithm}
\usepackage{xcolor}
\usepackage{multirow}
\usepackage{caption}
\usepackage{graphicx}
\usepackage{subfigure}
\usepackage{amsmath}
\usepackage{balance}
\usepackage{color}

\DeclareMathOperator{\E}{\mathbbm{E}}

\usepackage{tabularx}
\newcolumntype{C}[1]{>{\centering\arraybackslash}p{#1}}
\AtBeginDocument{%
  \providecommand\BibTeX{{%
    \normalfont B\kern-0.5em{\scshape i\kern-0.25em b}\kern-0.8em\TeX}}}

\copyrightyear{2024} 
\acmYear{2024} 
\setcopyright{acmlicensed}\acmConference[KDD '24]{Proceedings of the 30th ACM SIGKDD Conference on Knowledge Discovery and Data Mining}{August 25--29, 2024}{Barcelona, Spain}
\acmBooktitle{Proceedings of the 30th ACM SIGKDD Conference on Knowledge Discovery and Data Mining (KDD '24), August 25--29, 2024, Barcelona, Spain}
\acmDOI{10.1145/3637528.3671570}
\acmISBN{979-8-4007-0490-1/24/08}
\begin{document}



\title{A Self-boosted Framework for Calibrated Ranking}



\author{Shunyu Zhang}
\orcid{0009-0000-1936-1162}
\affiliation{%
  \institution{Kuaishou Technology}
  \city{Beijing}
  \country{China}}
\email{zhangshunyu@kuaishou.com}

\author{Hu Liu}
\orcid{0000-0003-2225-7387}
\authornote{Corresponding Author. }
\affiliation{%
  \institution{Kuaishou Technology}
  \city{Beijing}
  \country{China}}
\email{hooglecrystal@126.com}

\author{Wentian Bao}
\orcid{0009-0001-2195-0553}
\affiliation{%
  \institution{Columbia University}
  \city{Beijing}
  \country{China}}
\email{wb2328@columbia.edu}

\author{Enyun Yu}
\orcid{0009-0009-0847-7464}
\affiliation{%
  \institution{Northeasten University}
  \city{Beijing}
  \country{China}}
\email{yuenyun@126.com}

\author{Yang Song}
\orcid{0000-0002-1714-5527}
\affiliation{%
  \institution{Kuaishou Technology}
  \city{Beijing}
  \country{China}}
\email{yangsong@kuaishou.com}
\renewcommand{\shortauthors}{Shunyu Zhang, Hu Liu, Wentian Bao, Enyun Yu, \& Yang Song}

\renewcommand{\shortauthors}{Zhang and Liu, et al.}


\begin{abstract}

Scale-calibrated ranking systems are ubiquitous in real-world applications nowadays, which pursue accurate ranking quality and calibrated probabilistic predictions simultaneously. 
For instance, in the advertising ranking system, the predicted click-through rate (CTR) is utilized for ranking and required to be calibrated for the downstream cost-per-click ads bidding.
Recently, multi-objective based methods have been wildly adopted as a standard approach for Calibrated Ranking, which incorporates the combination of two loss functions: a pointwise loss that focuses on calibrated absolute values and a ranking loss that emphasizes relative orderings.
However, when applied to industrial online applications, existing multi-objective CR approaches still suffer from two crucial limitations. 
First, previous methods need to aggregate the full candidate list within a single mini-batch to compute the ranking loss. Such aggregation strategy violates extensive data shuffling which has long been proven beneficial for preventing overfitting, and thus
degrades the training effectiveness. 
Second, existing multi-objective methods apply the two inherently conflicting loss functions on a single probabilistic prediction, which results in a sub-optimal trade-off between calibration and ranking.

To tackle the two limitations, we propose a Self-Boosted framework for Calibrated Ranking (SBCR). 
In SBCR, the predicted ranking scores by the online deployed model are dumped into context features.
With these additional context features, each single item can perceive the overall distribution of scores in the whole ranking list, so that the ranking loss can be constructed without the need for sample aggregation. 
As the deployed model is a few versions older than the training model, the dumped predictions reveal what was failed to learn and keep boosting the model to correct previously mis-predicted items. 
Moreover, a calibration module is introduced to decouple the point loss and ranking loss. The two losses are applied before and after the calibration module separately, which elegantly addresses the sub-optimal trade-off problem.
We conduct comprehensive experiments on industrial scale datasets and online A/B tests, demonstrating that SBCR can achieve advanced performance on both calibration and ranking. Our method has been deployed on the video search system of Kuaishou, and results in significant performance improvements on CTR and the total amount of time users spend on Kuaishou.

\end{abstract}



\settopmatter{printfolios=true}

\maketitle

\section{Introduction}

As one of the most popular video-sharing apps in China, \textit{Kuaishou} strongly relies on its leading personalized ranking system, which serves hundreds of millions of users with fingertip connection to billions of attractive videos. 
Once the ranking system receives a request from a user (aka. \textit{query} in literature),
it will predict a \textit{ranking score} for each of the retrieved candidate videos (aka. \textit{items, documents}).  These ranking scores are not only used in sorting the candidate items to fit the user’s personalized interest, but also essential for many downstream applications. For example, we use the click-through rate (CTR) to guide ads bidding and the probability of effectively watching to estimate the video quality. 
This suggests that industrial ranking systems should emphasize two matters simultaneously:
1). the \textit{relative orders} between scores, namely the ranking quality  evaluated by GAUC and NDCG \cite{jarvelin2002cumulated}, and 2). the accurate \textit{absolute values} of scores which should be calibrated to some actual likelihood when mapped to probabilistic predictions.

To meet this practical demand in industrial ranking systems, there have been emerging studies on a paradigm known as Calibrated Ranking (CR) \cite{bai2023listce,yan2022scale}. The standard approach of CR usually incorporates a multi-objective loss function: a pointwise loss that focuses on calibrated absolute values and a ranking loss that emphasizes relative orderings. 
Sculley~\cite{sculley2010combined} combines regression and pairwise loss for CTR prediction with offline experiments. Further studies~\cite{bai2023listce,sheng2023jrc} combine pointwise loss and listwise loss for CR in real-world systems. 
Although encouraging progress has been made on both calibration and ranking ability, existing multi-objective CR approaches still suffer from two crucial limitations in industrial online applications. 

First, existing multi-objective CR approaches usually contradict extensive data shuffling which has long been proven beneficial in preventing the overfitting issue and is almost the default setting in industrial sequential gradient descent (Fig \ref{fig:shuffle}). Specifically, one component of the multi-objective loss, the ranking loss, is defined to make comparisons between scores of candidate items retrieved for the same request. This naturally requires the aggregation of the whole candidate list in a single mini-batch, making extensive item-level data shuffling inapplicable. 
As the result, the training process suffers from many terrible issues that could otherwise be avoided by extensive shuffling, including
the non-IID data, and overfitting caused by the aggregation of similar samples.
We will demonstrate the degradation of training effectiveness from insufficient shuffling in our experiments.

Second, conventional multi-objective CR applies point loss and ranking loss jointly on a single probabilistic prediction. However, the two objectives are not necessarily compatible, 
or even inherently conflicting, making the best trade-off sub-optimal for both calibration and ranking. And this trade-off is also sensitively dependent on the relative weights of the two objectives, leading to a challenging hyper-parameter choosing issue. How to decouple the two objects, namely optimizing one without sacrificing the other, still remains an open question. 

To address the first limitation, we proposed a self-boosted pairwise loss that enables extensive data shuffling, while achieving high ranking quality like what conventional ranking loss does. Our strategy is to dump the ranking scores predicted by the deployed model on our online server
into the context features of a specific query. With these contexts and negligible additional cost (only a few real numbers), each single candidate item can perceive the overall distribution of ranking scores under the same query. So there is no need to aggregate the whole candidate list in a single mini-batch for score comparison.
More importantly, as the deployed model is a few versions older than the training model, the dumped scores actually reveal what the model
failed to learn and further direct the following update to pay extra
attention to previously mispredicted items. We term this \textit{Self-Boosted}.
We further tackle the second limitation by introducing a calibration module that decouples the ranking and calibration losses. Ranking and calibration losses are applied before and after the calibration module separately, which elegantly addresses the sub-optimal multi-objective trade-off problem.

Finally, we propose the Self-Boosted framework for Calibrated Ranking (SBCR). Our architecture includes two modules, 1). a ranking module termed Self-Boosted Ranking (SBR) trained by a multi-objective loss consisting of the pointwise and proposed self-boosted pairwise losses, and 2). a following calibration module trained by a calibration loss. Our main contributions are summarized as follows:

\begin{figure}[!t]
    \centering
    \includegraphics[width=0.98\linewidth]{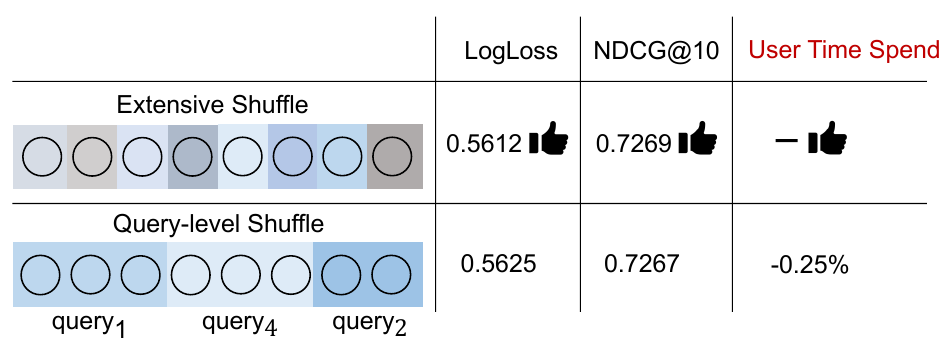}
    \caption[]{The performance comparison of different data shuffling strategies. 
    Evaluation metrics include: Logloss, NDCG@10 (widely used in ranking), and the total amount of time users spend on Kuaishou (the most important metric in our online A/B test).
    Extensive item-level data shuffling (upper) significantly outperforms the query-level data shuffling (bottom) where the whole candidate item list retrieved for a single request is aggregated in a single mini-batch. 
    Theoretical explanation will be discussed in Sec \ref{sec:limit}.
    This experimental result validates the advantage of extensive data shuffling and motivates us to propose a novel ranking loss that enables extensive shuffling. 
    \protect\footnotemark}
    \label{fig:shuffle}
\end{figure}
\footnotetext{The experiments are conducted on the pointwise production baseline with standard Logloss. The training details are described in Sec.\ref{exp:setup}}

\begin{itemize}

\item 
We highlight the two limitations of conventional multi-objective CR approaches in industrial online applications, namely, the contradiction with extensive data shuffling and the sub-optimal trade-off between calibration and ranking.

\item 
We propose a novel SBCR framework that successfully addresses the two limitations. In SBCR, ranking quality is emphasized without the need for data aggregating, and the two objectives are decoupled to avoid the conflict.

\item 
We validate SBCR on the video search system of Kuaishou.
Extensive offline experiments show that our method can achieve advanced performance on both calibration and ranking. 
In online A/B tests, SBCR also outperforms the strong production baseline and brings significant improvements on CTR and the total amount of time users spend on Kuaishou. SBCR has now been deployed on the video search system of Kuaishou, serving the main traffic of hundreds of millions of active users.

\end{itemize}

\section{Related Work}
We mainly focus on the related work concerning these aspects: CTR Prediction, Learning-to-Rank (LTR), and Calibrated Ranking.

\textbf{CTR} Prediction aims to predict a user’s personalized click tendency, which is crucial for nowadays information systems.
In the last decades, the field of CTR prediction has evolved from traditional shallow models, e.g. Logistic Regression (LR), Gradient Boosting Decision Tree (GBDT)~\cite{ke2017lightgbm}, to deep neural models e.g. Wide \& Deep~\cite{cheng2016wide}, DCN~\cite{wang2017deep}. 
Most researches are dedicated to improving model architectures: Wide \& Deep~\cite{cheng2016wide} and DeepFM~\cite{guo2017deepfm} combine low-order and high-order features to improve model expressiveness, and DCN~\cite{wang2017deep,wang2021dcn} replace FM of DeepFM with Cross Network. DIN~\cite{zhou2018din} and SIM~\cite{pi2020sim} employ the attention mechanism to extract user interest.
Despite recent progress, the loss function is still not well-explored and the dominant pointwise LogLoss~\cite{zhou2018din} can't well satisfy the ranking quality highly desired in practice~\cite{li2015click,sculley2010combined,liu2009learning}.

\textbf{Learning-To-Rank (LTR)} generally learns a scoring function to predict and sort a list of objects~\cite{liu2009learning,XiaLWZL2008ListMLE,bruch2019analysis}. The evaluation is based on ranking metrics considering the sorted order, such as Area Under the Curve~\cite{Fawcett2006ROC} and Normalized Discounted Cumulative Gain~\cite{jarvelin2002cumulated}. 
The pointwise approach~\cite{gey1994inferring} learns
from the label of a single item. And the pairwise methods learn from the relative ordering of item pairs~\cite{BurgesSRLDHH2005RankNet,GuSTRL2015RankingSVM}, which is further used in real-world LTR systems~\cite{burges2010ranknet,huang2020embedding}.
Some others propose the ranking-metric-based optimization including the listwise approaches, which directly target on aligning the loss with the evaluation metrics~\cite{cao2007learning,qin2010general}. However, the poor calibration ability limits non-pointwise LTRs for wider applications. 

\textbf{Multi-Objective CR} is a natural idea to address the above problems, where ranking loss is calibrated by a point loss~\cite{li2015click,yan2022scale} to preserve both calibration and ranking ability. 
\citet{sculley2010combined} conducts an early study to combine regression with pairwise ranking, and \citet{li2015click} shows it can yield promising results in real-world CTR prediction systems. Recently, multi-objective CR has been deployed in deep models and more methods are proposed to better combine the two losses. \citet{yan2022scale} address the training divergence issues and achieve calibrated outputs. \citet{bai2023listce} further proposes a regression-compatible ranking approach to balance the calibration and ranking accuracy. \citet{sheng2023jrc} propose a hybrid method using two logits corresponding to click and non-click states, to jointly optimize the ranking and calibration.
However, existing multi-objective methods are still sub-optimal facing the trade-off between calibration and ranking, and contradict extensive data shuffling as we have stated.

Another line of works on \textbf{CR} revolves around post-processing methods including Platt-scaling, Isotonic Regression, and etc. \citet{tagami2013ctr} adopt pairwise squared hinge loss for training, and then used Platt-scaling \cite{1999platt} to convert the ranking scores to probabilities. \citet{chaudhuri2017ranking} compare post-processing methods including Platt-scaling and Isotonic Regression, to calibrate the outputs of an ordinal regression model. Apart from the calibration tailored for ranking, there are some others that only aim for more accurate probabilistic predictions, including binning~\cite{naeini2015obtaining} and hybrid approaches~\cite{kumar2019verified}, for e.g., Smooth Isotonic Regression~\cite{deng2021calibrating}, Neural Calibration~\cite{pan2020field}. While they all require extra post-processing, our method is jointly learned during training.

\section{Methodology}\label{section3}
We start from reviewing the general preliminaries of CR in Section \ref{section3.1}. Then we describe its standard approach multi-objective CR and analyze the two main drawbacks in Section \ref{multi-obj}. To address these issues, we finally dig into the details of our proposed SBCR in Section \ref{CSP sec}. The notations used are summarized in Table \ref{notations}.

\subsection{Preliminaries}\label{section3.1}

\begin{table}
\caption{Important Notations Used in Section 3.}\label{notations}
\begin{center}
\begin{tabular}{ll|ll|ll}
   \toprule
$\mathcal Q$ & query feature space		&	$q$ &query 			&$\hat y$ & predicted score	\\
$\mathcal X$ & item feature space & $x$ & item & $\mathcal Y$ & label space \\
$s$ & score function & $\sigma$ & sigmoid & $\mathbb R$ & real number\\
    
\midrule
\end{tabular}
\begin{tabular}{ll|ll}
$\mathcal D$ & training dataset& $y$ & groundtruth label \\
$\mathcal L_{point}$& pointwise loss & $\ell_{point}$ & point loss for one sample\\
$\mathcal L_{pair}$ & pairwise loss & $\ell_{pair}$ & pair loss for one query\\
$Q$ & \# queries &$\alpha$, $\beta$ & trade-off parameter  \\
$\mathcal D_q^+$ & positive item set & $\mathcal D_q^-$& negative item set\\
$\mathbb I$& indicator & $\mathcal L_{multi}$ & multi-objective loss \\
$\widetilde{\mathbf s_q}$& dumped scores &$\widetilde{\mathbf y_q}$& dumped labels\\
$k$& interval index &$\widetilde{s}$ & deployed score function \\
$\mathbf a$ & interval heights &$g$ & calibration module \\
$\mathcal L_{cali}$ & calibration loss & 
$\hat y_{cali}$& calibrated prediction \\

\midrule
\end{tabular}\end{center}
\begin{tabular}{ll}
$b_k$ & the function value of an interval end  \qquad \qquad    \qquad   \\
$\mathcal L_{pair\_boost}$& self boost loss\\
$\ell_{pair\_boost}$ & self boost loss for one sample \\
\bottomrule
\end{tabular}
\end{table}

The aim of \textit{Calibrated Ranking} \cite{bai2023listce,yan2022scale} is to predict ranking 
scores for a list of candidate \textit{items} (or documents, objects) properly given a certain \textit{query} (request with user, context features). 
Besides the \textit{relative order} between the ranking scores emphasized by conventional LTR \cite{cao2007learning}, CR also focuses on the \textit{calibrated absolute values} of scores simultaneously.
To be specific, ranking scores should not only improve ranking metrics, such as GAUC and NDCG \cite{jarvelin2002cumulated}, but also be scale-calibrated to some actual likelihood when mapped to probabilistic predictions.

Formally, our
goal is to learn a scoring function
$s: \mathcal Q \times \mathcal X \rightarrow \mathbb R$, given a training dataset $\mathcal D$ consisting of a list of samples $(q, x, y)$. Here, we denote $\mathcal Q$ as the query space, 
$q\in\mathcal Q$ as a query , $\mathcal X$ as the item feature space, $x\in \mathcal X$ as a candidate item, and
$y\in \mathcal Y$ as a ground-truth label under specific business settings. For example, in our video search system at Kuaishou, $y$ can indicate whether a video is clicked, liked, or watched effectively (beyond a predefined time threshold).
Without loss of generality, we assume the label space $\mathcal Y=\{1, 0\}$ throughout this paper.

The model first predicts a ranking 
score (also known as \textit{logit}),
\begin{equation}
s_{q,x} = s(q, x),
\end{equation}
for each item $x$ associated with the same query $q$, and then ranks the items according to the descending order of the scores. In addition,
the ranking score is mapped to a probabilistic prediction by a sigmoid function, 
\begin{equation}
\hat y_{q,x} = \sigma(s_{q,x}),
\end{equation}
which will be used for many downstream applications. 
And this probabilistic prediction should be well calibrated, i.e., agree with the actual likelihood that the item is clicked, liked, or watched effectively, i.e.,
$\hat y_{q,x} = \mathbb E[y|q,x]$.

\subsection{Multi-objective Calibrated Ranking}\label{multi-obj}
In literature and industrial ranking systems, multi-objective methods have been wildly adopted as standard approaches for CR \cite{bai2023listce}. 
\subsubsection{Existing Methods}
The key idea of multi-objective CR is to take the advantage of two loss functions: 1). a pointwise loss that calibrates the absolute probabilistic prediction values, and 2). a ranking loss that emphasizes the relative orders of items associated with the same query.

Specifically, the pointwise loss is defined as the average of Cross Entropy overall training samples,
\begin{equation} 
\label{eq:point}
\begin{aligned}
     \mathcal{L}_{point} =&-\frac{1}{|\mathcal D|}\sum_{(q,x,y)\in \mathcal D} \ell_{point}(q,x,y)\\
     =&-\frac{1}{|\mathcal D|}\sum_{(q,x,y)\in \mathcal D} y \log \hat y_{q,x} + (1 - y) \log(1 - \hat y_{q,x}),
\end{aligned}
\end{equation}
which is shown to be calibrated \cite{yan2022scale} since the minima is achieved at the point 
$
    \hat y_{q,x} = \E[y | q, x]
$.

The ranking loss is usually defined on pairs of training samples. We denote $\mathcal D_q^+ = \{x|(q,x,y)\in \mathcal D, y=1\}$, $\mathcal D_q^- = \{x|(q,x,y)\in \mathcal D, y=0\}$ as the set of positive and negative items associate to the same query $q$. And the pairwise loss is defined to promote a large margin between items across the two sets,
 
\begin{equation}
\label{eq:pair}
\begin{aligned}
\mathcal{L}_{pair} =&-\frac{1}{Q} \sum_{q\in \mathcal D} \ell_{pair} (q)\\
=& -\frac{1}{Q} \sum_{q\in \mathcal D}\frac{1}{|\mathcal D_q^+||\mathcal D_q^-|} \sum_{i\in \mathcal D_q^+, j\in\mathcal D_q^-}\log \sigma(s_{q,i} - s_{q, j}),
\end{aligned}
\end{equation}
where the outer average is taken over all queries in $\mathcal D$ \footnote{We slightly abuse the notation for conciseness of Eq. \ref{eq:pair}: $q\in \mathcal D$ represents any unique query in the training dataset, namely, $q\in\{q|(q,x,y)\in \mathcal D\}$.}, $Q$ denotes the number of unique queries, and
the inner average is taken over pairs inside each query. 
Although achieving high ranking quality, this pairwise loss
suffers from the miscalibration problem due to translation-invariant \cite{yan2022scale}. 

To combine the advantages of both, a multi-objective loss is defined as
\begin{equation}
\mathcal L_{multi} = \alpha \mathcal L_{point} + (1-\alpha) \mathcal L_{pair},
\label{eq:multi}
\end{equation}
where $\alpha\in (0,1)$ controls the trade-off between the quality of calibration and ranking.

Note that besides the pairwise loss, there are many 
other widely used ranking losses, such as listwise softmax \cite{bruch2019analysis,pasumarthi2019tf} and listwise ApproxNDCG~\cite{qin2010general}. Similarly, these listwise losses also sacrifice calibration for ranking performance, and could be used as a component in multi-objective CR. We skip the discussion for conciseness.

\subsubsection{Limitations of Existing Multi-Objective CR}
\label{sec:limit}
Despite being extensively studied, existing multi-objective CR approaches still suffer from two crucial limitations:  contradiction to extensive data shuffling and sub-optimal trade-off between calibration and ranking abilities. 

In the default setting of most industrial training systems, samples are first \textit{extensively} shuffled, and then divided into mini-batches for sequential gradient descent. 
This shuffling simulates the Independent and Identically Distribution (IID), and consequently prevents the overfitting problem caused by aggregated similar samples inside each mini-batch. 
We show the performance gain from data shuffling in our experiments in Fig. \ref{fig:shuffle}. 

While in multi-objective CR, extensive data shuffling is not applicable due to the definition of the pairwise loss. Specifically, to calculate
$\ell_{pair}(q)$, we need to go through all positive-negative pairs associated with $q$, indicating that all samples with the same query have to be gathered inside a single mini-batch. 
As a result, in training pairwise (or listwise) losses, data can only be shuffled at query level, not sample level.
This aggregation of similar samples 
(under the same query, with identical context, user features)
inside each mini-batch contradicts the IID assumption and thus heavily degrades the performance gain from data shuffling. 

Another limitation of conventional CR lies in the trade-off nature of the multi-objective loss. To be specific, we introduce the following theorem.

\noindent
\begin{theorem}
$\mathcal L_{pair}$ and $\mathcal L_{point}$ have distinct optimal solutions.
\end{theorem}

\noindent
\begin{proof}
As mentioned in Sec 3 of \cite{bai2023listce},
$\mathcal L_{point}$ is minimized when, 
\begin{equation}\label{proof_1}
\sigma(s_{q,x}) = \mathbb{E}[y|q,x].
\end{equation}
And $\mathcal L_{pair}$ is minimized when,
\begin{equation}\label{proof_2}
\sigma(s_{q,x_1}-s_{q,x_2}) = \mathbb{E}[\mathbb{I}(y_1>y_2)|q,x_1,x_2].
\end{equation}
In the case $y_1$ and $y_2$ are independent, we rewrite Eq. \ref{proof_2} as,
\begin{equation}\label{proof_3}
\sigma(s_{q,x_1}-s_{q,x_2}) =\mathbb{E}[y_1|q,x_1]*(1-\mathbb{E}[y_2|q,x_2]).
\end{equation}
Supposing the two losses share the same optimal solution, we use Eq. \ref{proof_1} to further rewrite Eq. \ref{proof_3} as,
\begin{equation*}
\sigma(s_{q,x_1}-s_{q,x_2})=\sigma(s_{q,x_1})*(1-\sigma(s_{q,x_2})).
\end{equation*}
Thus,
\begin{equation*}
\frac{1}{1+ e^{s_{q,x_2}-s_{q,x_1}}}=\frac{1}{1+ e^{s_{q,x_2}-s_{q,x_1}} + e^{-s_{q,x_1}} + e^{s_{q,x_2}}}.
\end{equation*}
Ultimately, we have derived an infeasible equation, $e^{-s_{q,x_1}} + e^{s_{q,x_2}}=0$, indicating that the two losses have distinct optimal solution.

\end{proof}

Intuitively, $\mathcal L_{pair}$ emphasizes the relative order of items, while failing to predict the absolute probabilistic value accurately. And $\mathcal L_{point}$ vice versa. 
In Eq \ref{eq:multi}, however, the two inherently conflicting losses are applied on the prediction $s_{q,x}$, and thus the best trade-off may be sub-optimal for both calibration and ranking.
In addition, this trade-off also sensitively depends on $\alpha$, leading to a challenging hyper-parameters choosing issue.
In contrast, a better design would 
decouple the ranking and calibration losses with separated network structures and gradient cut-off strategies, which 
elegantly addresses the objective conflict.


\begin{figure*}[!ht]
    \centering
    \includegraphics[width=0.98\linewidth]{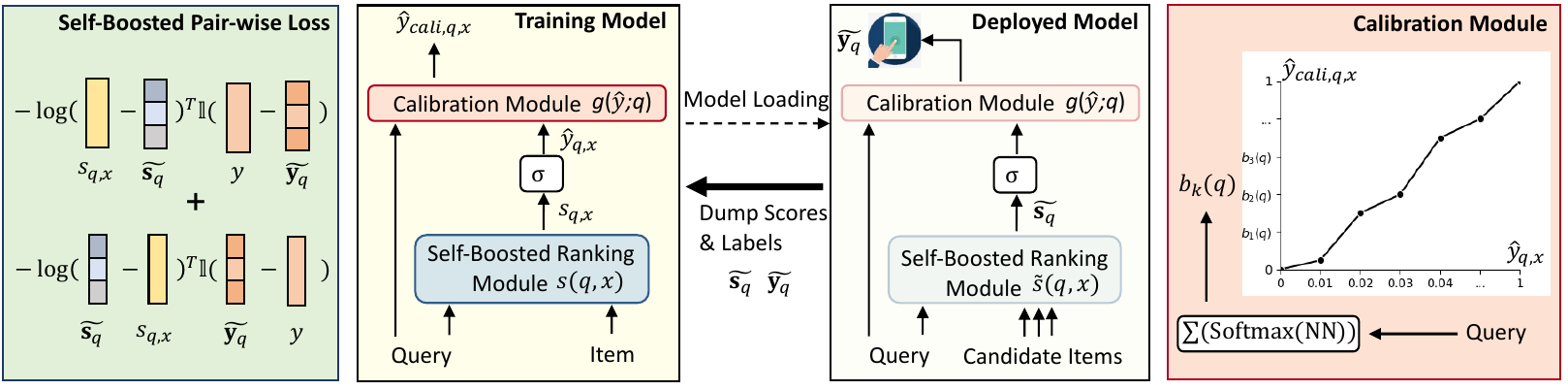}
    \caption{The architecture of the proposed Self-Boosted framework for Calibrated Ranking. 
    Middle: SBCR consists of two modules: a self-boosted ranking module (SBR)  trained by a multi-objective loss (pointwise and self-boosted pairwise loss) and a calibration module.
    Left: the details of the proposed self-boosted pairwise loss. Using dumped ranking scores $\widetilde{\mathbf s_q}$ from the online deployed model, we enable both comparisons between samples associated with the same query and extensive sample-level data shuffling.
    Right: the proposed calibration module that decouples the ranking and calibration objectives to avoid the conflict.}
    \label{fig:model}
\end{figure*}

\subsection{Self-Boosted Calibrated Ranking}
\label{CSP sec}
To address the two limitations, our proposed SBCR mainly consists of two modules: 
1). a self-boosted ranking module (SBR) that enables extensive data shuffling, while achieving high ranking quality like what conventional pairwise loss does, and 2). an auxiliary calibration module that decouples the ranking and calibration losses and thus successfully addresses the 
sub-optimal trade-off problem.
We describe our model architecture in Fig. \ref{fig:model}.

\subsubsection{The Self-Boosted Ranking Module}
As mentioned earlier, $\ell_{pair}(q)$ is calculated on all possible positive-negative item pairs associated with $q$, making it inseparable for sample-level data shuffling. To address this issue, we propose a novel loss function,
\begin{equation}
\mathcal L_{pair\_boost} = -\frac{1}{|\mathcal D|} \sum_{(q,x,y)\in\mathcal D} \ell_{pair\_boost}(q,x,y).
\end{equation}
Different from $\mathcal L_{pair}$ (Eq. \ref{eq:pair}), here each component $\ell_{pair\_boost}$ only depends on a single training sample.
Note this design is nontrivial due to the conflict between two facts: 
\begin{itemize}
    \item In order to enhance the ranking ability, comparisons between the current item and its peers, i.e., other items associated with the same query, are essential for ranking order learning. So
    $\ell_{pair\_boost}$ should be able to perceive the overall score distribution under $q$. 
    
    \item 
    Since only one item's feature $x$ is used as the input, it is impossible to run the network forward and backward for the other items associated with $q$.
\end{itemize}
We now dig into the details of $\ell_{pair\_boost}$ that solves this conflict. 

Our system mainly consists of two parts: 1). an online \textit{Server} that receives a user's request, makes real-time responses by scoring and ranking candidate items, and dumps logs including the user's feedback. and 2). a near-line \textit{Trainer} that sequentially trains the scoring function $s(\cdot,\cdot)$ on latest logs.
For every few minutes, 
the Server continuously loads the latest scoring function from the Trainer. 
We denote the deployed model on the Server as $\widetilde s(\cdot,\cdot)$, which 
would be a bit older than the training model $s(\cdot,\cdot)$. 

Formally, when the Server receives 
a query $q$ and candidates $[x_1,..., x_n]$, it 
\begin{enumerate}
    \item first predicts ranking scores 
$\widetilde{\mathbf s_q} = [\widetilde s_{q, x_1},..., \widetilde s_{q,x_n}]$ for candidates using the deployed model $\widetilde s(\cdot,\cdot)$, 
    \item then presents the ranked items to the user and collects her feedback,
$\widetilde {\mathbf y_q} = [\widetilde y_{q,x_1},...,\widetilde y_{q,x_n}]$.
    \item finally dumps the scores $\widetilde{\mathbf s_q}\in \mathbb R^ n$ and labels $\widetilde {\mathbf y_q} \in \mathbb R ^n$ into context features of the current query $q$. 
\end{enumerate}
Although only adding negligible cost, 
$\widetilde{\mathbf s_q}$ and $\widetilde {\mathbf y_q}$ actually provide rich knowledge on the score distribution to enhance the model's ranking ability.
More importantly, the dumped scores from an older model reveal what the Trainer failed to learn and further direct the following update to pay extra attention to the previously mis-predicted samples. We thus term this \textit{Self-Boosted}. 

With the extra $\widetilde{\mathbf s_q}$ and $\widetilde{\mathbf y_q}$ as context features in $q$,
we define,
\begin{equation} 
\label{eq:pairboost}
\begin{aligned}
\ell_{pair\_boost}(q,x,y) = &-\log \sigma(s_{q,x} -\widetilde{\mathbf s_q}) ^\top \mathbb I (y-\widetilde{\mathbf y_q})\\
&
-\log \sigma(\widetilde{\mathbf s_q}-s_{q,x}) ^\top \mathbb I (\widetilde{\mathbf y_q}-y),
\end{aligned}\end{equation}
where the indicator $\mathbb I(z) = 1$ if $z>0$ and 0 otherwise. We slightly abuse notations by broadcasting scalars to fit vectors' dimensions. 

\textbf{Remarks}: Similar to $\ell_{pair}$, our $\ell_{pair\_boost}$ also compares items under the same query and encourages large margins. 
It is different that the self-boost mechanism 
keeps lifting the model performance by focusing on previously missed knowledge.
We take the first term in $\ell_{pair\_boost}$ for example. After masked by the indicator, it promotes the margins between the positive item's score $s_{q,x}$ and its negative peers' dumped scores $\widetilde{\mathbf s_q}$. If the items under $q$ were poorly predicted previously, the unmasked elements in $\widetilde{\mathbf s_q}$ would be larger, resulting in a larger loss value. To achieve a satisfactory margin, $s_{q,x}$ would be lifted more aggressively in the backpropagation.

\subsubsection{The Calibration Module}
The probabilistic prediction $\hat y$ trained using
multi-objective CR (Eq. \ref{eq:multi}) usually suffers from the sub-optimal trade-off between calibration and ranking ability. 

To address this issue, we propose a calibration module to decouple the two losses.
The pairwise loss and point loss are applied
before and after the calibration module separately, which elegantly addresses the objective conflicting problem. 

The calibration module maps $\hat y$ into a calibrated one:
\begin{equation}\label{eq:cali}
\hat y_\text{cali} = g(\hat y;q),
\end{equation}
where $\hat y_\text{cali} \in (0,1)$ is the calibrated probability and $g$ is the proposed calibration module. 
We make several considerations in the design of $g(\cdot)$:
    
    First, to be flexible enough to capture various functional distributions, $g(\cdot)$ is set as a continuous piece-wise linear function. Without loss of generality, we partition the function domain $(0,1)$ into 100 equal-width intervals and set $g(\cdot)$ to be a linear function inside each interval: 
    \begin{equation}\begin{aligned}
    \hat y_\text{cali} 
    =\sum_{k =0}^{99}\left[b_k + (\hat y - 0.01*k)\frac{b_{k+1}-b_{k}}{0.01} \right] \mathbb I_{k}(\hat y),
    \end{aligned}\label{eq:cali_interval}
    \end{equation}
    where $\mathbb I_k(\hat y)$ indicates whether $\hat y$ lies inside the $k$-th interval. Namely $\mathbb I_k(\hat y) = 1$ if $\hat y \in [0.01*k, 0.01*(k+1))$ and 0 otherwise.
    And $b_k, k \in\{0,...,100\}$ are the function values at all interval ends. Obviously, $b_0 = 0$ and $b_{100} = 1$. And the other 99 which control the function property will be adjusted during modeling training.
    
    Second, to keep the knowledge learnt previously from pairwise loss, our calibration module should preserve the relative orders of items under the same query. Namely, given any two predictions $\hat y_{q, i} < \hat y_{q, j}$, the calibrated probability $\hat y_{\text{cali},q,i} \leq \hat y_{\text{cali},q,j}$. We thus require the piece-wise linear function to be non-decreasing, i.e., 
    $b_0 \leq b_1 \leq...\leq b_{100}$. And the parameters $b_1,...,b_{99}$ are learnt \textit{only} from the features of the current query, such as the user features (user id, gender, age, behavior sequences) and context features (timestamp, page index, date). So Eq. \ref{eq:cali} only includes $q$ as the input but excludes any item features in $x$.
    The learning process is defined:
    \begin{equation}\begin{aligned}\label{calib_a}
    b_k(q) = \sum_{j=1}^k a_j, k \in \{1,...,100\}, \\
    \mathbf a = \text{Softmax}(\text{NeurNet}(q)).
    \end{aligned}\end{equation}
    where $\mathbf a\in \mathbb R^{100}$ represents the learnt 100 interval heights, normalized by the softmax function and $b_k$ is calculated as the accumulated heights from the first to $k$-th interval.

Finally, our calibration module is trained using pointwise loss to ensure accurate absolute prediction values:
\begin{equation}\label{calib_final}
\mathcal{L}_{cali} = -\frac{1}{|\mathcal D|}\sum_{(q,x,y)\in \mathcal D} \left[y\log\hat y_{\text{cali},q,x} + (1-y)\log(1-\hat y_{\text{cali},q,x})\right].
\end{equation}


\subsubsection{The Overall Architecture of SBCR and Training Tricks}
Our model consists of two networks, as summarized in Fig \ref{fig:model}. 

The first deep network defines the scoring function $s(q,x)$ with two groups of inputs: 1). $q$, features shared by samples under the same query, including user features, long-term user behaviors and context features; 2). $x$, features of a specific item. 
In the industrial ranking system of Kuaishou, we adopt QIN~\cite{guo2023query} as $s(q,x)$ 
due to its SOTA performance. 
This network is trained using,
\begin{equation}
\label{eq:multiboost}
\mathcal L_{multi\_boost} = \alpha \mathcal L_{point} + (1-\alpha) \mathcal L_{pair\_boost}.
\end{equation}
We replace the conventional pairwise loss in 
Eq. \ref{eq:multi} by our proposed self-boosted pairwise loss, which enables sample-level shuffling. 

Our second deep network defines the function for calculating the interval heights $\mathbf a$ in Eq. \ref{calib_a}, which is trained by Eq. \ref{calib_final}. In our system, the network consists of 4 FC layers of size $(255, 127, 127,100)$.

To avoid a sub-optimal trade-off between calibration and ranking, we further stop the gradient back propagation for all inputs to the calibration module (i.e., $\hat y, q$). Thus $s(q,x)$ focuses on the ranking quality and the calibration module only deals with calibration. We will discuss more on parameter sensitivity in our experiments.

\section{Experiments}
In this section, to validate the effectiveness of our proposed SBCR framework, we compare SBCR with many state-of-the-art CR algorithms. 
We also provide an in-depth analysis to investigate the impact of each building block in SBCR.
Experiments are conducted based on the Kuaishou video search system, including both offline evaluations on the billion-scale production dataset, and online A/B testing on the real traffic of millions of active users. We did not include experiments on the public datasets, since our method is designed for the online training system.


\subsection{Experiment Setup}
\label{exp:setup}

\subsubsection{Datasets. }
In our offline experiments, all compared algorithms are initialized from the same checkpoint, trained online using 5 days' data, and then frozen to test on the 6th day's data. 
The dataset is collected from the user log on the video search system of Kuaishou and the statistics of the dataset are shown in Tab. \ref{tab:dataset}.
Our method is designed for the online training systems that are widely deployed in industrial scenarios, so we only conduct experiments on the real production dataset.

\begin{table}[!tbp]
\centering
\caption{Statistics of the real production dataset. Query means a request from the user. M and B are short for million and billion.}
\label{tab:dataset}
    \begin{tabular}{c|cccc}
    \toprule
     Dataset      &  \#Querys   &  \#Items  &   \#Samples & \#Samples/\#Query   \\
    \midrule
    Train   & 211 M  & 452 M  & 1.88 B. & 8.89 \\
    Test   & 41.2 M &  281 M & 	372 M &  9.04 \\
    \bottomrule
    \end{tabular}
\end{table}

\begin{table*}[!t]
\centering
\caption{The comparison of SOTAs on the production dataset. The best results are highlighted in bold and the second-best are underlined. * indicates significant improvement with p-value=0.05 than previous best-performing Point + ListCE. Shuffle denotes compatibility with extensive sample-level data shuffling. SBCR outperforms all due to 2 key advantages: compatibility with extensive data 
shuffling and effective structure to decouple ranking and calibration losses.}
\label{tab:main}
\resizebox{0.98\textwidth}{!}
{
\begin{tabular}{@{}cc|C{1.2cm}c|C{1.2cm}C{1.2cm}C{1.2cm}@{}|c}
\toprule
\multicolumn{2}{c|}{\multirow{2}{*}{Method}} & \multicolumn{2}{c|}{Ranking Metrics} & \multicolumn{3}{c|}{Calibration Metrics} &\multirow{2}{*}{Sample-level Shuffle}\\ \cmidrule(l){3-7} 
\multicolumn{2}{c|}{} & GAUC  & NDCG@10 & LogLoss & PCOC & ECE \\ \midrule
\multicolumn{1}{c|}{\multirow{4}{*}{Single-objective}} & Pointwise & 0.6076 & 0.7269 & \underline{0.5612} & 1.0183 & 0.0047 & \checkmark\\
\multicolumn{1}{c|}{} & RankNet~\cite{burges2010ranknet} & \underline{0.6110} & 0.7297 & 2.7337 & 7.8285 & 0.5438 &$\times$\\
\multicolumn{1}{c|}{} & ListNet~\cite{cao2007learning} & 
0.6107 & \underline{0.7303} & 2.9845 & 8.1322 & 0.5975 &$\times$\\
\multicolumn{1}{c|}{} & ListCE~\cite{bai2023listce} & 0.6092 & 0.7285 & 0.5838 & 1.0420 & 0.0092 &$\times$\\ \midrule
\multicolumn{1}{c|}{\multirow{4}{*}{Multi-objective}} 
& Point + RankNet~\cite{li2015click} & 0.6089 & 0.7279 & 0.5623 & 0.9759 & 0.0057&$\times$ \\
\multicolumn{1}{c|}{} & Point + ListNet~\cite{yan2022scale} & 0.6095 & 0.7282 & 0.5621 & 1.0293 & 0.0065 &$\times$\\
\multicolumn{1}{c|}{} & Multi-task~\cite{yan2022scale} &  0.6082 & 0.7276 & 0.5615 &  1.0207 & 0.0052  &$\times$\\
\multicolumn{1}{c|}{} & Calibrated Softmax~\cite{yan2022scale} &  0.6105 & 0.7302 & 0.5678 & 1.1532 &  0.0126 &$\times$\\
\multicolumn{1}{c|}{} & JRC~\cite{sheng2023jrc} & 0.6102 & 0.7293 & 0.5619 & \underline{0.9894} & 0.0049 &$\times$\\
\multicolumn{1}{c|}{} & Point + ListCE~\cite{bai2023listce} & 0.6107 & 0.7298 & 0.5615 & 0.9878 & \textbf{0.0044}&$\times$\\
\midrule
\multicolumn{1}{c|}{Ours} & SBCR & \textbf{0.6118*}  & \textbf{0.7315*}  & \textbf{0.5610*}  & \textbf{1.0092*}  & \underline{0.0045} &\checkmark\\ 
\bottomrule
\end{tabular}
}
\end{table*}

\subsubsection{Implementation Details.}

In our experiments, we adopt QIN~\cite{guo2023query} as our architecture for all compared methods.
With efficient user behavior modeling, QIN is a strong production baseline latest deployed on the KuaiShou video search system.
For feature engineering, ID features are converted to dense embeddings and concatenated with numerical features. All models are trained and tested using the same optimization settings, i.e., Adam optimizer, learning rate of 0.001, and batch size of 512. 
All models are trained with one epoch following \citet{ZhangSZJHDZ2022OneEpoch}, which is widely adopted in the production practice. 
For the relative ranking weight $(1-\alpha)/\alpha$, we chose the best one from [0.01, 0.1, 1.0, 10, 100] for each compared algorithm and report the performance of them in their own optimal hyper-parameter settings 
for a fair comparison. For our proposed SBCR, we simply set the relative weight to 1 consistently across all experiments.

\subsubsection{Evaluation Metrics.}

In this work, we consider both ranking and calibration performances. For evaluating the ranking performance, we choose \textbf{NDCG@10} and \textbf{GAUC} (Group AUC). NDCG@10 is consistent with other metrics like NDCG@5.
GAUC is widely employed to assess the ranking performance of items associated with the same query, and has demonstrated consistent with online performance in previous studies~\cite{zhou2018din,pi2020sim}. GAUC is computed by:
\begin{equation}
\label{eqn:gauc}
\begin{aligned}
    \textrm{GAUC} = & \frac{\sum_{q\in \mathcal D} \# \textrm{candidates}(q) \times \textrm{AUC}_q}
{\sum_{q\in \mathcal D} \# \textrm{candidates}(q)}, \\
\end{aligned}
\end{equation}
where $\textrm{AUC}_q$ represent AUC within the same query $q$.

For evaluating the calibration performance, we include \textbf{LogLoss}, expected calibration error (\textbf{ECE})~\cite{pan2020field}, and predicted CTR over the true CTR (\textbf{PCOC})~\cite{GuoPSW2017OnCalibration}. LogLoss is calculated the same way as in Eq. \ref{eq:point}, which measures the logarithmic loss between probabilistic predictions and true labels. 
ECE and PCOC are computed by 
\begin{equation}
\label{eqn:calmetric}
\begin{aligned}
    \text{ECE} & = \frac{1}{|\mathcal D|}\sum_{k=0}^{99} |\sum_{(q,x,y)\in \mathcal D} (y-\hat y_{\text{cali},q,x})\mathbb I_{k}(\hat y_{\text{cali},q,x})|, \\
    \text{PCOC} &  = \frac{1}{\sum_{(q,x,y)\in \mathcal D} y}\sum_{(q,x,y)\in \mathcal D} \hat y_{\text{cali},q,x}, \\
\end{aligned}
\end{equation}
where $\mathbb I_{k}$ is defined in the same way as Eq. \ref{eq:cali_interval}. 
Among the three calibration metrics, 
LogLoss provides sample-level measurement, whereas ECE and PCOC provide subset level and dataset level measurement, respectively. There are some other metrics like Cal-N and GC-N~\cite{deng2021calibrating} and here we mainly follow the setting of calibrated ranking~\cite{yan2022scale,sheng2023jrc} for a fair comparison. Lower LogLoss or ECE indicates better performance, and PCOC is desired to be close to 1.0.

These five metrics serve as reliable indicators for evaluating both ranking and calibration abilities. 

\subsubsection{Compared Methods}
As in Table \ref{tab:main}, we include several important baseline methods for a comprehensive comparison. These baseline methods are divided into two groups based on whether the loss function is single-objective or multi-objective. The \textbf{Single-objective} group consists of these four methods:
\begin{itemize}
    \item \textbf{Pointwise} refers to the standard LogLoss (Eq. \ref{eq:point}), which is widely adopted for binary targets. It is also the production baseline for most industrial ranking systems. 
    \item \textbf{RankNet}~\cite{burges2010ranknet} adopts pairwise loss (Eq. \ref{eq:pair}) to optimize the relative ranking of pairs of samples.
    \item \textbf{ListNet}~\cite{cao2007learning} defines a listwise loss to maximize the likelihood of the correct ordering of the whole list. 
    \item \textbf{ListCE}~\cite{bai2023listce} proposes a regression compatible ranking approach where the two ranking and regression components are mutually aligned in a modified listwise loss. 
\end{itemize} 

In the \textbf{Multi-objective} group, we include several advanced methods with both calibrated point loss and ranking-oriented loss for the comprehensive comparison:
\begin{itemize}
    \item \textbf{Point + RankNet}~\cite{li2015click} combines the pointwise and pairwise 
 loss in a multi-objective paradigm (Eq. \ref{eq:multi}) to improve both calibration and ranking.
    \item \textbf{Point + ListNet}~\cite{yan2022scale} is the combination of the pointwise and the listwise loss, which is proved as a strong calibrated ranking baseline and termed ``multi-ojbective" in ~\cite{yan2022scale}. 
    \item \textbf{Multi-task}~\cite{yan2022scale} is a multi-task method that uses multi-head of DNN for the ranking and calibration scores.
    \item \textbf{Calibrated Softmax}~\cite{yan2022scale} is a reference-based method where an anchor candidate with label $y_0$ to a query is introduced to control the trade-off.

    \item \textbf{JRC}~\cite{sheng2023jrc} proposes a hybrid method that employs two logits corresponding to click and non-click states and jointly optimizes the ranking and calibration abilities.
    \item \textbf{Point + ListCE}~\cite{bai2023listce} combines the pointwise loss with ListCE that makes ranking and regression components compatible and achieves advanced results.

\end{itemize}

\subsection{Main Experimental Results}
The main experimental results are shown in Table \ref{tab:main}.
The methods are compared on both ranking and calibration metrics.
From the results, we have the following observations:

First, in the single-objective group, pointwise achieves the best calibration performance but inferior ranking performance. In contrast, RankNet and ListNet outperform the Pointwise model on the ranking ability, at the expense of being completely uncalibrated. Among these methods, ListCE can reach a tradeoff for it makes regression compatible with ranking, but it still suffers from poor calibration. This validates the necessity of calibrated ranking.

Second, the multi-objective methods incorporate the two losses and achieve a better trade-off. And several recent studies (JRC and Point + ListCE) have achieved encouraging progress on both ranking and calibration metrics. Note that LogLoss is a stronger calibration metric than PCOC and ECE. For example, if a model predicts averaged prediction over the whole dataset, the model achieves perfect PCOC and ECE but poor LogLoss. Considering LogLoss as the main calibration metric, all compared multi-objective methods still suffer from sub-optimal trade-off between calibration and ranking, when compared to the best Single-objective methods, validating our second motivation. When comparing the three variants of \cite{yan2022scale}, we find that cal-softmax achieves slightly better NDCG but suffers from unacceptable worst calibration performance, while multi-objective gets the best trade-off between ranking and calibration. Our observation are also consistent with the previous results reported in the original paper.

Third, our proposed SBCR outperforms all compared algorithms on both ranking and calibration performance. This validates the two key advantages of SBCR, i.e., compatibility with extensive data-shuffling and effective structure to avoid trade-off between ranking and calibration. Specifically, SBCR is trained with the dumped predictions by the online deployed model, making it the only method that needs no sample aggregation when computing the ranking loss. And a calibration module is introduced to decouple the point loss and ranking loss to address the sub-optimal trade-off problem. The gain of the two key advantages will be further analyzed in section \ref{sec:abla_shuffle} and \ref{sec:abla_cali}.

\begin{table}[!tbp]
\centering
\caption{Comparison on different sample aggregations. \textit{Shuf} denotes extensive sample-level shuffle, \textit{Aggre} means aggregating the whole candidate list in a single mini-batch. Best are marked bold. \textit{Shuf} consistently outperforms \textit{Aggre} on both Point and Point + Pair models.}
\label{tab:expshuffle}
\resizebox{0.46\textwidth}{!}
{
    \begin{tabular}{l|l|cccc}
    \toprule
       &  & {NDCG@10} & {LogLoss} & {PCOC} & {ECE}\\
    \midrule	
\multirow{2}{*}{Point} & Aggre & 0.7267 & 0.5625 & 1.0205 & 0.0049 \\
      & Shuf & 0.7269  & \textbf{0.5612}  & 1.0183  & \textbf{0.0046}  \\ 
\midrule
Point  & Aggre& 0.7279 & 0.5614 & 0.9759 & 0.0057\\
      + Pair       & Shuf (SBR)& \textbf{0.7315} & 0.5614 & \textbf{1.0147} & 0.0056\\
    \bottomrule
    \end{tabular}
}
\end{table}

\subsection{Ablation Study and Analysis}
We conduct ablation studies to investigate the contribution of each SBCR building block: the self-boosted pairwise loss to address the data-shuffling problem, and the calibration module to address the trade-off problem. We also include hyper-parameter analysis to show the sensitivity.

\subsubsection{Analysis on Data-shuffling}
\label{sec:abla_shuffle}
As mentioned, extensive data-shuffling that simulates the Independent and Identical Distribution has long been proven beneficial
in preventing the overfitting problem. We re-validate the performance gain of data-shuffling on both point loss and point + pair loss.
For point + pair loss, Point + RankNet~\cite{li2015click} is used as the aggregation method and our SBR (Eq. \ref{eq:multiboost}) is used as the shuffling method, results shown in Tab. \ref{tab:expshuffle}.

Shuffling achieves consistent improvement over sample aggregation, which validates that 
extensive data-shuffling is essential for performance and supports our first motivation. Conventional multi-objective CR algorithms require the aggregation of the whole candidate list in a single mini-batch for computing the ranking loss, which is incompatible with extensive data-shuffling. SBCR solves this problem and achieves superior performance.

\begin{table}[!tbp]
\centering
\caption{Comparison on different Calibration Modules. ``Calib'' is our proposed one.
We observe consistent improvement when the ranking module is trained by ListNet and SBR, validating our adaptability.}
\label{tab:expcalib}
{
    \begin{tabular}{l|l|ccccc}
    \toprule
     &     & {GAUC} & {LogLoss} & {PCOC} & {ECE}\\
    \midrule	
\multirow{3}{*}{ListNet}  & $\varnothing$ & 0.6107 & 2.9845 & 8.1322 & 0.5975 \\
  & Platt & 0.6107 & 0.5627 & 1.0353 & 0.0073\\
  & Calib & 0.6107 & 0.5621 & 1.0279 & 0.0065 \\
  \midrule
\multirow{3}{*}{SBR}  & $\varnothing$ & \textbf{0.6118} & 0.5614  & 1.0147 & 0.0056 \\
  & Platt  & \textbf{0.6118} & 0.5613  & 1.0132 & 0.0051 \\
  & Calib & \textbf{0.6118}& \textbf{0.5610} &\textbf{1.0092} & \textbf{0.0045}  \\
    \bottomrule
    \end{tabular}
}
\end{table}

\subsubsection{The Impact of Calibration Module}
\label{sec:abla_cali}
In order to analyze the impact of the Calibration Module, we compared several methods in Tab. \ref{tab:expcalib}. 
ListNet-Platt~\cite{yan2022scale} applies Platt-scaling post-processing after a ListNet model, which is a strong baseline. Hence, we compared our calibration module with Platt-scaling on the same ListNet model. 
We also apply Platt-scaling and ours upon the same SBR model as defined in Eq. \ref{eq:multiboost}.

As shown in Tab. \ref{tab:expcalib}, we observed that both Platt-scaling and ours preserve the relative orders and show the same GAUC. Our proposed calibration module achieves consistently better calibration ability on both ListNet and SBR, which validates the strong adaptability of our method.

\subsubsection{Effects of Hyper-Parameter}
\begin{figure}[!t]
    \centering
    \includegraphics[width=0.82\linewidth]{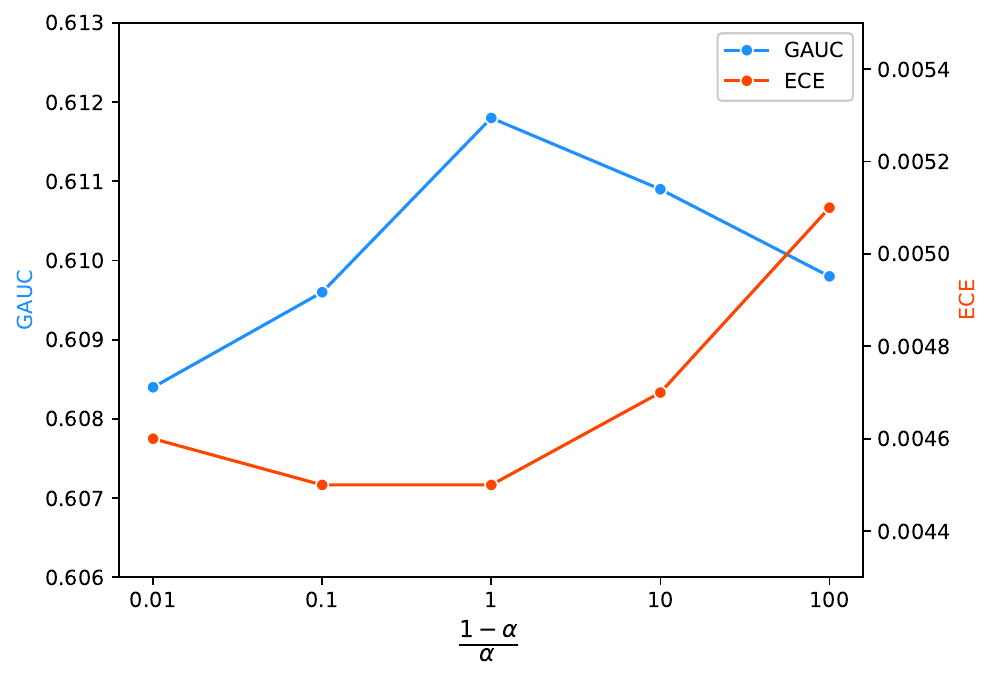}
    \caption{The sensitivity of relative ranking weight $(1-\alpha)/\alpha$ for SBCR (Eq. \ref{eq:multiboost}). Higher GAUC and lower ECE indicate better performance.}
    \label{fig:sensitivity}
\end{figure}

The only hyper-parameter introduced in our method is $\alpha$, the trade-off parameter between point loss and self-boosted pair loss in Eq. \ref{eq:multiboost}. We define $(1-\alpha)/\alpha$ as the relative ranking weight and examine the sensitivity in Fig. \ref{fig:sensitivity}.

First, surprisingly, an extremely small value of relative ranking weight is not optimal for calibration and an extremely large value is not optimal for ranking. 
This validates that the two losses collaborate with each other. Point loss is necessary for ranking, especially in queries where all items are negative and pair loss is necessary for calibration by giving auxiliary guild for model training.

Second, SBCR is robust to the setting of $\alpha$. When the relative weight is set 10 times larger /smaller, the performance of SBCR is still comparable to that of SOTA.
We own this robustness to our calibration module which is specially designed to be monotonic for a given $q$. Namely, it preserves the learned orderings from our ranking module. Optimizing the calibration module will not degrade the ranking performance.

Third, we still observe slightly trade-off between calibration and ranking. This trade-off has been greatly improved by the calibration module.
We tried to remove the calibration module and found that when $(1-\alpha)/\alpha=100$, ECE will be increased to 0.1796, which is 35 times worse than the current case.

\subsection{Online Performance}
We validate the proposed SBCR with online A/B testing on the video search system of Kuaishou (Tab. \ref{tab:online}).
We compare SBCR with our latest production baseline \textit{Point} and the strongest compared algorithm reported in Table \ref{tab:main}, \textit{Point+ListCE}.
All three algorithms share the same backbone
QIN~\cite{guo2023query}, and the same features, model structures and optimizer.
The online evaluation metrics are CTR, View Count and User Time Spend. View count measures the total amount of video users watched, and time spend measures the total amount of time users spend on viewing the videos.

As shown in Tab. \ref{tab:online}, SBCR contributes to the +4.81\% increase in CTR, +3.15\% increase in View Count and +0.85\% increase in Time Spend. Note that the 0.2\% increase is a significant improvement in our system. SBCR is also efficient for online serving. Compared to our baseline QIN+Point, the only additional module is the calibration network that adds 1.41\% parameters and 0.96\% floating point operations, which is negligible for the model.

Note that there is an important issue that is commonly faced in the real production systems.
Usually, there are several different algorithms under A/B test 
simultaneously. So the dumped scores used for training SBCR are actually from deployed models in different A/B tests, not only SBCR's corresponding deployed model. This is not equivalent to the standard self-boosted mechanism in Sec 3. We should check the impact.

In A/B test, the pointwise baseline serves the main traffic, thus SBCR is mostly trained with dumped scores from the pointwise baseline. In this sub-optimal implementation, we still observe significant gain (the 3rd two in Table \ref{tab:online}).
And in A/B backtest (last row in Table \ref{tab:online}), SBCR serves the main traffic and the old pointwise model serves the small traffic. The dumped scores are mostly from SBCR itself. We observe a larger improvement compared to that in AB test: an additional gain of +0.63\% View Count and +0.43\% Time Spent. We conclude that the dumped scores from other models also work, with a slightly smaller gain compared to standard self-boost. This is because the mis-predicted samples by other models are also hard and informative and focusing on other strong model's mistakes is also beneficial.

\begin{table}[!tbp]
\centering
\caption{The improvements of SBCR in online A/B test compared to the production baseline. In the video search system of Kuaishou, 0.2\% increase is a significant improvement for CTR, View Count and User Time spend.}
\label{tab:online}
    \begin{tabular}{l|ccc}
    \toprule
           &  CTR  & View Count &  Time Spent  \\
    \midrule
    QIN + Point (baseline) & - & - & - \\
    QIN + Point + ListCE  & +1.01\% & +0.80\% & +0.14\% \\
    QIN + SBCR (AB)  &  +4.81\% & +3.15\% & +0.85\%  \\
    QIN + SBCR (AB back) & +5.70\% & +3.78\%& +1.28\% \\
    \bottomrule
    \end{tabular}
\end{table}

\section{Conclusion and Future Works}
We proposed a Self-Boosted framework for Calibrated Ranking in industrial online applications. SBCR addressed the two limitations of conventional multi-objective CR, namely, the contradiction with extensive data shuffling and the sub-optimal trade-off between calibration and ranking, which contributes to significant performance gain. SBCR outperformed our highly optimized production baseline and has been deployed on the video search system of Kuaishou, serving the main traffic of hundreds of millions of active users.

Note that we restricted our calibration module $g(\cdot)$ to piece-wise linear function since it is easy to guarantee the monotonicity, which is necessary to preserve the relative orders of items under the same query. A promising future direction is to improve the flexibility of $g$ by upgrading it to monotonic neural networks.



\bibliographystyle{ACM-Reference-Format}
\balance
\bibliography{sample-base}

\appendix




\end{document}